\documentclass[12pt]{article}
\usepackage{graphicx}
\usepackage{latexsym,amsmath,amsfonts,amssymb}
\newcommand{\be}{\begin{equation}}
\newcommand{\ee}{\end{equation}}
\newcommand{\beq}{\begin{equation}}
\newcommand{\eeq}{\end{equation}}
\newcommand{\bea}{\begin{eqnarray}}
\newcommand{\eea}{\end{eqnarray}}

\newcommand{\ba}{\begin{eqnarray}}
\newcommand{\ea}{\end{eqnarray}}
\begin{document}
\baselineskip=15.5pt
\pagestyle{plain}
\setcounter{page}{1}


\def\del{{\partial}}
\def\vev#1{\left\langle #1 \right\rangle}
\def\cn{{\cal N}}
\def\co{{\cal O}}
\def\IC{{\mathbb C}}
\def\IR{{\mathbb R}}
\def\IZ{{\mathbb Z}}
\def\RP{{\bf RP}}
\def\CP{{\bf CP}}
\def\Poincare{{Poincar\'e }}
\def\tr{{\rm tr}}
\def\tp{{\tilde \Phi}}

\def\TL{\hfil$\displaystyle{##}$}
\def\TR{$\displaystyle{{}##}$\hfil}
\def\TC{\hfil$\displaystyle{##}$\hfil}
\def\TT{\hbox{##}}
\def\HLINE{\noalign{\vskip1\jot}\hline\noalign{\vskip1\jot}}
\def\seqalign#1#2{\vcenter{\openup1\jot
  \halign{\strut #1\cr #2 \cr}}}
\def\lbldef#1#2{\expandafter\gdef\csname #1\endcsname {#2}}
\def\eqn#1#2{\lbldef{#1}{(\ref{#1})}%
\begin{equation} #2 \label{#1} \end{equation}}
\def\eqalign#1{\vcenter{\openup1\jot
    \halign{\strut\span\TL & \span\TR\cr #1 \cr
   }}}
\def\eno#1{(\ref{#1})}
\def\href#1#2{#2}
\def\half{{1 \over 2}}

\def\ads{{\it AdS}}
\def\adsp{{\it AdS}$_{p+2}$}
\def\cft{{\it CFT}}
\def\dx{{\partial_x}}
\def\dph{{\partial_{\phi_1}}}
\def\ders{{\partial_\sigma}}

\newcommand{\ber}{\begin{eqnarray}}
\newcommand{\eer}{\end{eqnarray}}

\newcommand{\beqar}{\begin{eqnarray}}
\newcommand{\cN}{{\cal N}}
\newcommand{\cO}{{\cal O}}
\newcommand{\cA}{{\cal A}}
\newcommand{\cT}{{\cal T}}
\newcommand{\cF}{{\cal F}}
\newcommand{\cC}{{\cal C}}
\newcommand{\cR}{{\cal R}}
\newcommand{\cW}{{\cal W}}
\newcommand{\eeqar}{\end{eqnarray}}
\newcommand{\tht}{\thteta}
\newcommand{\lm}{\lambda}\newcommand{\Lm}{\Lambda}
\newcommand{\eps}{\epsilon}


\newcommand{\nonu}{\nonumber}
\newcommand{\oh}{\displaystyle{\frac{1}{2}}}
\newcommand{\dsl}
  {\kern.06em\hbox{\raise.15ex\hbox{$/$}\kern-.56em\hbox{$\partial$}}}
\newcommand{\id}{i\!\!\not\!\partial}
\newcommand{\as}{\not\!\! A}
\newcommand{\ps}{\not\! p}
\newcommand{\ks}{\not\! k}
\newcommand{\D}{{\cal{D}}}
\newcommand{\dv}{d^2x}
\newcommand{\Z}{{\cal Z}}
\newcommand{\N}{{\cal N}}
\newcommand{\Dsl}{\not\!\! D}
\newcommand{\Bsl}{\not\!\! B}
\newcommand{\Psl}{\not\!\! P}
\newcommand{\eeqarr}{\end{eqnarray}}
\newcommand{\ZZ}{{\rm \kern 0.275em Z \kern -0.92em Z}\;}

                                                                                                    
\def\del{{\delta^{\hbox{\sevenrm B}}}} \def\ex{{\hbox{\rm e}}}
\def\azb{A_{\bar z}} \def\az{A_z} \def\bzb{B_{\bar z}} \def\bz{B_z}
\def\czb{C_{\bar z}} \def\cz{C_z} \def\dzb{D_{\bar z}} \def\dz{D_z}
\def\im{{\hbox{\rm Im}}} \def\mod{{\hbox{\rm mod}}} \def\tr{{\hbox{\rm Tr}}}
\def\ch{{\hbox{\rm ch}}} \def\imp{{\hbox{\sevenrm Im}}}
\def\trp{{\hbox{\sevenrm Tr}}} \def\vol{{\hbox{\rm Vol}}}
\def\rl{\Lambda_{\hbox{\sevenrm R}}} \def\wl{\Lambda_{\hbox{\sevenrm W}}}
\def\fc{{\cal F}_{k+\cox}} \def\vev{vacuum expectation value}
\def\nodiv{\mid{\hbox{\hskip-7.8pt/}}}
\def\ie{{\em i.e.}}
\def\ie{\hbox{\it i.e.}}

\def\CC{{\mathchoice
{\rm C\mkern-8mu\vrule height1.45ex depth-.05ex
width.05em\mkern9mu\kern-.05em}
{\rm C\mkern-8mu\vrule height1.45ex depth-.05ex
width.05em\mkern9mu\kern-.05em}
{\rm C\mkern-8mu\vrule height1ex depth-.07ex
width.035em\mkern9mu\kern-.035em}
{\rm C\mkern-8mu\vrule height.65ex depth-.1ex
width.025em\mkern8mu\kern-.025em}}}
                                                                                                    
\def\RR{{\rm I\kern-1.6pt {\rm R}}}
\def\NN{{\rm I\!N}}
\def\ZZ{{\rm Z}\kern-3.8pt {\rm Z} \kern2pt}
\def\IB{\relax{\rm I\kern-.18em B}}
\def\ID{\relax{\rm I\kern-.18em D}}
\def\II{\relax{\rm I\kern-.18em I}}
\def\IP{\relax{\rm I\kern-.18em P}}
\newcommand{\CS}{{\scriptstyle {\rm CS}}}
\newcommand{\CSs}{{\scriptscriptstyle {\rm CS}}}
\newcommand{\rc}{\nonumber\\}
\newcommand{\bear}{\begin{eqnarray}}
\newcommand{\eear}{\end{eqnarray}}
\newcommand{\W}{{\cal W}}
\newcommand{\F}{{\cal F}}
\newcommand{\x}{{\cal O}}
\newcommand{\LL}{{\cal L}}
                                                                                                    
\def\mani{{\cal M}}
\def\calo{{\cal O}}
\def\calb{{\cal B}}
\def\calw{{\cal W}}
\def\calz{{\cal Z}}
\def\cald{{\cal D}}
\def\calc{{\cal C}}
\def\to{\rightarrow}
\def\ele{{\hbox{\sevenrm L}}}
\def\ere{{\hbox{\sevenrm R}}}
\def\zb{{\bar z}}
\def\wb{{\bar w}}
\def\nodiv{\mid{\hbox{\hskip-7.8pt/}}}
\def\menos{\hbox{\hskip-2.9pt}}
\def\dr{\dot R_}
\def\drr{\dot r_}
\def\ds{\dot s_}
\def\da{\dot A_}
\def\dga{\dot \gamma_}
\def\ga{\gamma_}
\def\G{\Gamma}
\def\dal{\dot\alpha_}
\def\al{\alpha_}
\def\cl{{closed}}
\def\cls{{closing}}
\def\vev{vacuum expectation value}
\def\tr{{\rm Tr}}
\def\to{\rightarrow}
\def\too{\longrightarrow}


\def\a{\alpha}
\def\b{\beta}
\def\c{\gamma}
\def\d{\delta}
\def\e{\epsilon}           
\def\f{\phi}               
\def\vf{\varphi}  \def\tvf{\tilde{\varphi}}
\def\vp{\varphi}
\def\g{\gamma}
\def\h{\eta}
\def\i{\iota}
\def\j{\psi}
\def\k{\kappa}                    
\def\l{\lambda}
\def\m{\mu}
\def\n{\nu}
\def\o{\omega}  \def\w{\omega}
\def\q{\theta}  \def\th{\theta}                  
\def\r{\rho}                                     
\def\s{\sigma}                                   
\def\t{\tau}
\def\u{\upsilon}
\def\x{\xi}
\def\z{\zeta}
\def\pt{\tilde{\varphi}}
\def\tt{\tilde{\theta}}
\def\lab{\label}  
\def\6{\partial}
\def\wg{\wedge}
\def\atanh{{\rm arctanh}}
\def\bpsi{\bar{\psi}}
\def\bt{\bar{\theta}}
\def\bvf{\bar{\varphi}}

%
                                                                                                    
\newfont{\namefont}{cmr10}
\newfont{\addfont}{cmti7 scaled 1440}
\newfont{\boldmathfont}{cmbx10}
\newfont{\headfontb}{cmbx10 scaled 1728}
\renewcommand{\theequation}{{\rm\thesection.\arabic{equation}}}
\begin{titlepage}

\begin{center} \Large \bf On unquenched ${\cal 
N}=2$ holographic flavor

\end{center}

\vskip 0.3truein
\begin{center}

\'Angel 
Paredes\footnote{angel.paredes@cpht.polytechnique.fr}
\vspace{0.7in}\\
 \it{Centre de Physique Th\'eorique\\ \'Ecole Polytechnique\\
and UMR du CNRS 7644\\
91128 Palaiseau, France}
\vspace{0.3in}
\end{center}
\vskip 1cm
\centerline{{\bf Abstract}}
\vspace{0.1in}
The addition of fundamental degrees of freedom to a theory which
is dual (at low energies) to ${\cal N}=2$ SYM in 1+3 dimensions
is studied. The gauge theory
lives on a stack of $N_c$ D5 branes wrapping an $S^2$ with the appropriate twist,
while the fundamental hypermultiplets are introduced by adding a different
set of $N_f$ D5-branes. In a simple case, a system of first order equations taking into account
the backreaction of the $N_f \sim N_c$ flavor branes is derived.
From it, the modification of the holomorphic coupling is computed explicitly.
Mesonic excitations are also discussed.

\vskip1truecm
\vspace{0.3in}
\leftline{CPHT-RR 080.1006}
\leftline{hep-th/0610270}
\smallskip
\end{titlepage}

\section{Introduction and summary of results}
\label{intro}
\setcounter{equation}{0}

In its original and most studied form \cite{Maldacena:1997re}, the gauge-string
correspondence relates a field theory where all fields transform in the adjoint 
representation of the gauge group to a theory of closed strings. An enormous
amount of work has
been devoted to generalize the correspondence in different directions. Since the physics
of real world strong interactions is dominated by mesons and baryons, a
generalization of obvious importance is to include matter transforming in the fundamental
representation of the gauge group. Early works on the subject were 
\cite{Grana:2001xn,Bertolini:2001qa,Karch:2000gx}.

In the seminal papers \cite{Karch:2002sh,Kruczenski:2003be}, 
it was argued that including
a small number of flavors ($N_f$ fixed, $N_c\to\infty$) corresponds in the dual 
theory to adding an open
string sector due to the presence of a few flavor branes. 
In the field theory, this is a quenched
approximation in the sense that the diagrams with fundamentals running in the loops
are suppressed by $N_c^{-1}$. In the gravity side, one can neglect the
backreaction of the branes on the geometry. These ideas and  methods sparked a lot
of activity which led to a neat understanding of many different phenomena in such a 
limit, in a variety of setups.

Nevertheless, field theories in which $N_f$ is of the same order as $N_c$ are
amazingly rich as is clear, for instance, from Seiberg's analysis of ${\cal N}=1$ SQCD.
In addition, the fact that in the real world $\frac{N_f}{N_c}$ is of order one is reflected
in experimental facts as, among others, multihadron production or the large mass of the
$\eta'$ meson. This motivates the study of string duals in the Veneziano limit
$N_c \to \infty$ with $g^2 N_c$ and $\frac{N_f}{N_c}$ fixed. In order to maintain the
curvature small, strong coupling (large $g^2 N_c$) is required.
However, this case is not as well understood as the quenched one. Some
works along this direction studying four-dimensional field theories are
\cite{Grana:2001xn,Bertolini:2001qa,Bertolini:2002xu,Nastase:2003dd,Burrington:2004id,Kirsch:2005uy,
Casero:2006pt} (three-dimensional field theories were addressed in \cite{Cherkis:2002ir,
Erdmenger:2004dk,Gomez-Reino:2004pw}). In one way or another, all these models have
a region of large curvature associated to the presence of the fundamental quarks.

In this note, a setup consisting of wrapped branes, dual to ${\cal N}=2$ SQCD is presented.
By appropriately smearing the flavor branes such that the shell of branes
forms a domain wall in the geometry,
they do not generate any pathological region in the geometry
(as long as $N_f\leq 2N_c$ so there is no Landau pole in the dual field theory).
This smearing corresponds to a smearing of the
complex masses of the hypermultiplets, more details are given
in section \ref{sect:back}.
Anyhow,
the solution has a naked curvature singularity
in the IR and a divergent dilaton in the UV. These features are already present in the original setup
and are not related to the presence of flavor branes. Some comments on how these pathologies
can be dealt with can be found in sections \ref{review} and \ref{sect:mesons} respectively.

The framework is the gravity dual of ${\cal N}=2$ SYM introduced in 
\cite{Gauntlett:2001ps,Bigazzi:2001aj}. It consists of D5-branes 
wrapping an $S^2$ inside a $CY_2$ in such
a way that eight supercharges are preserved. At small energies compared to the inverse radius
of the sphere, the theory becomes effectively four-dimensional. It will be shown how one
can add a different kind of D5-branes in a way that no further supersymmetry is broken.
They introduce an open string sector that corresponds to having in the field theory (massive) hypermultiplets
transforming in the fundamental representation of the gauge group. As explained above, when
the number of these branes is comparable to the number of colors, their backreaction on the
geometry has to be taken into account. This is done by applying the procedure advocated in
\cite{Casero:2006pt}: the D5-branes are sources for the Ramond-Ramond $F_{(3)}$ field
strength, and therefore they
modify its Bianchi identity. Once this is taken into account, the set of first order equations
can be found by imposing the vanishing of the supersymmetry variation of the fermion fields.
The resulting system of equations is quite complicated and I have not managed to find an
analytical solution. However, as it will be shown, it can indeed be explicitly solved in a certain
region of the space. This is enough to prove that the modification of the couplings due to
the flavor branes is the one expected from field theory. 
Finally, the meson spectrum is studied. 
The excitations can be rearranged in a tower
of massive ${\cal N}=2$ vector multiplets. Due to the 
UV pathological behavior of the
solution, a regularization procedure is needed in order 
to estimate the masses of the physical excitations.
Once this procedure is implemented, the dependence of the 
meson masses on the different physical parameters is briefly discussed.

Apart from the interest of the particular model that 
will be addressed, one of the aims of 
this paper is to make clearer the methods of 
\cite{Casero:2006pt} by their application
to a simpler setup. In particular, as stated above, 
in this ${\cal N}=2$ construction the flavor branes
do not spark the divergence or vanishing of any component of the metric.
Hopefully, this note can contribute in the development of techniques
for the study of holographic duals with unquenched matter.

The contents of the paper are organized as follows: in section \ref{review},
the unflavored solution is reviewed and rewritten with a notation that will be useful in the
following. In section \ref{flavorprobes}, the embedding of the flavor branes is discussed and
the physical meaning of the different spacetime directions is explained.
In section \ref{sect:back}, the system of equations including the backreaction is obtained
and in section \ref{sect:features} it is shown how the couplings are modified by the
unquenched flavor. Section 
\ref{sect:mesons} deals with the spectrum of excitations.
A few WKB formulae used in section \ref{sect:estim} are summarized
in appendix \ref{appendix}.

\section{A review of the dual of ${\cal N}=2$ SYM}
\label{review}
\setcounter{equation}{0}

In references \cite{Gauntlett:2001ps,Bigazzi:2001aj},
a IIB gravity dual of ${\cal N}=2$ SYM in the Coulomb branch was
found. The ten-dimensional solution was obtained as an uplift from
seven-dimensional $SO(4)$ gauged supergravity. It corresponds to 
5-branes\footnote{Originally, the solution was built with NS5-branes. 
In the following, the S-dual solution corresponding to D5-branes 
will be considered.} wrapping a two-sphere with the appropriate twisting
to preserve eight supercharges, {\it i.e.} ${\cal N}=2$ in the
effective four-dimensional low energy theory. Geometrically, it corresponds
to wrapping the branes along a compact SLag two-cycle inside a Calabi-Yau two-fold.
This leaves two flat transverse dimensions which are identified with
the moduli space corresponding to giving vevs to the complex scalar
inside the ${\cal N}=2$ vector multiplet.

The solution presented below has a curvature singularity in the IR.
The singularity is good according to the criterion of \cite{Maldacena:2000mw}.
In fact, it was shown in \cite{Hori:2002cd}, by considering wrapped NS5-branes
that the singularity is an artifact of the supergravity approximation and is
resolved by the worldsheet CFT. Therefore, the singularity will not be a matter of 
concern in the following.

This section does not contain original material
but reviews the results of 
 \cite{Gauntlett:2001ps,Bigazzi:2001aj,DiVecchia:2002ks} in
order to fix notation for the following 
and make this note reasonably self-contained.
An excellent review of this model and similar constructions is
\cite{Bigazzi:2003ui}.

\subsection{The solution}
\label{sect: gaun}

The metric (Einstein frame) of the solution found in \cite{Gauntlett:2001ps,Bigazzi:2001aj} reads:
\bear
ds_{10}^2&=&N_c g_s \a'e^\frac{\Phi}{2}\left[
\frac{1}{N_c g_s \a'} dx_{1,3}^2 +
z(d\tt^2 + \sin^2 \tt d\tvf^2) + e^{2x} dz^2 + d\th^2 +\right. \rc
&&\left. 
+ \frac{e^{-x}}{\Omega} \cos^2 \th (d\phi_1 + \cos\th d\tvf)^2 +
\frac{e^{x}}{\Omega} \sin^2 \th d\phi_2^2
\right]\,\,,
\label{metric1}
\eear
there is a magnetic RR twoform:
\beq
C_{(2)} = N_c g_s \a' \phi_2 d\left[ \frac{\sin^2 \th}{\Omega e^x}
(d\phi_1 +  \cos\tt d\tvf)\right]
\label{C21}
\eeq
and the dilaton is given by:
\beq
e^{2\Phi}=e^{2\Phi_0}e^{2z} \left[1-\sin^2 \th \frac {1+c e^{-2z}}{2z}\right]
\eeq
and we have defined the quantities:
\bear
e^{-2x}=1- \frac{1+ c e^{-2z}}{2z} \rc
\Omega = e^x \cos^2 \th + e^{-x} \sin^2 \th
\label{defs}
\eear
The angles take values in the following
intervals: $\tt \in [0,\pi],\ \th \in [0,\frac{\pi}{2}],\
\tvf,\phi_1,\phi_2 \in [0,2\pi)$. The variable $z$ ranges from
a value $z_0$ for which $e^{-2x(z_0)}=0$ up to infinity.

There are two integration constants on which the solution depends, 
$\Phi_0$ and $c$.
Following the discussion of \cite{Gauntlett:2001ps,Bigazzi:2001aj}, the 
solution is dual to ${\cal N}=2$
SYM at points of the Coulomb branch in which the vevs for 
the entries of the complex scalar
matrix
are distributed in a ring (they have fixed modulus whereas 
the phase is smeared). The constant
$c\geq -1$ determines the size of the ring. Solutions 
with $c<-1$ are unphysical and, in fact, the singularity 
becomes of the bad type.
Notice that although
taking a single vev different from zero
 spontaneously breaks the $U(1)_R$, this
smearing procedure restores it, at least if one ignores $N_c^{-1}$
effects, where one could start seeing that the distribution is not 
continuous (see figure \ref{picture} in section \ref{sect:back}).
Points of the Coulomb branch where the $U(1)_R$ is not restored were
discussed in \cite{Bigazzi:2001aj}.

The quantization condition is:
\beq
\frac{1}{2\kappa_{(10)}^2}\int_{S^3} F_{(3)} = N_c T_5
\label{quant}
\eeq
where $S^3$ is the transverse three-sphere parameterized by
$\th,\phi_1,\phi_2$ at $z\to \infty$, yielding:
\beq
\int_{S^3} F_{(3)} = N_c g_s \a' \int_0^{2\pi} d\phi_1
\int_0^{2\pi} d\phi_2 \int_0^{\frac{\pi}{2}} 2\sin\th\cos\th d\th=
4\pi^2 N_c g_s \a'
\label{quant2}
\eeq
Taking into account:
\beq
T_5=\frac{1}{(2\pi)^5 g_s \a'^3}\,,\qquad
\frac{1}{2\kappa_{(10)}^2}= \frac{1}{(2\pi)^7 g_s^2 \a'^4}
\label{t5k10}
\eeq
one can immediately check that (\ref{quant}) is satisfied.

\subsection{Rewriting the solution and Killing spinors}
\label{sect: solution}

In reference \cite{DiVecchia:2002ks}, the solution was rewritten
in different variables which allow a better understanding of the physics.
Define:
\beq
\rho = \sin\th e^z \,,\qquad \sigma = \sqrt{z} \cos\th e^{z-x}\,,
\label{varchange}
\eeq
so
the metric (\ref{metric1}) reads:
\bear
ds_{10}^2 &=& g_s N_c \a' e^\frac{\Phi}{2} \left[ 
\frac{1}{g_s N_c \a'} dx_{1,3}^2 +  z (d\tt^2 +
\sin^2 \tt d\tvf^2) + \right.\rc
&&
\left.
+ e^{-2\Phi} (d\rho^2 + \rho^2 d\phi_2^2)
+ \frac{e^{-2\Phi}}{z} \left( d\s^2 + \s^2
(d\phi_1 + \cos \tt d\tvf)^2 
\right) \right]
\label{metric2}
\eear
Written in this way, it is clear that the Calabi-Yau twofold directions are $\tt,\tvf,\s,\phi_1$
(of course, in this solution with fluxes there is not a 
Calabi-Yau any more, but it can be
thought of as a deformation of the Calabi-Yau that was present 
before backreaction).
The coordinates
$\rho,\phi_2$ 
span the transverse two-dimensional plane, so they should be identified with the moduli space,
and therefore rotations in $\phi_2$ are related to the $U(1)_R$ symmetry of the field theory.
These statements will be made more precise below.

Actually, one could  start by writing the metric (\ref{metric2})
as an ansatz with $z(\rho,\s)$, $\Phi(\rho,\s)$. The ansatz for
the RR 3-form would be:
\bear
F_{(3)}&=& N_c g_s \a' \left[ -g' d\phi_2 \wedge d\r \wedge
(d\phi_1 + \cos \tt d\tvf) - \dot g d\phi_2 \wedge d\s \wedge
(d\phi_1 + \cos \tt d\tvf)+\right. \rc
&& \left. + g \sin \tt d\phi_2\wedge d\tt \wedge d\tvf \right]
\label{F3}
\eear
which ensures $dF_{(3)}=0$ and we have defined:
\beq
' \equiv \partial_\rho \,,\qquad \dot{}  \equiv \partial_\s
\eeq
Introducing this ansatz in the IIB transformations for the fermions:
\bear
\d\l&=&\frac12 \partial_\mu \Phi \G^{\m} \s_1 \e +\frac{1}{24} e^\frac{\Phi}{2}
\G^{\m_1\m_2\m_3}\e F_{\m_1\m_2\m_3}=0\,,\rc
\d\psi_\m &=& \partial_\mu \e +\frac14 \omega_\m^{ab} \G^{ab} \e+
\frac{1}{96} e^\frac{\Phi}{2}\left(\G_\m^{\ \m_1\m_2\m_3}-9 \d_\m^{\m_1}
\G^{\m_2\m_3}\right) \s_1 \e  F_{\m_1\m_2\m_3}=0\,,
\label{susytransf}
\eear
a set of first order equations can be obtained. 
Consider the orthonormal frame:
\bear
&&e^i = e^\frac{\Phi}{4} dx_i\  (i=0,...,3)\,, \quad
e^4 = \sqrt{g_s N_c \a'} e^\frac{\Phi}{4} \sqrt{z} d\tt 
\,, \quad e^5  =  \sqrt{g_s N_c \a'} e^\frac{\Phi}{4} \sqrt{z}
\sin \tt d\tvf \,,\rc
&&e^6 = \sqrt{g_s N_c \a'} e^{-\frac{3\Phi}{4}}d\r \,, \quad
e^7 = \sqrt{g_s N_c \a'} e^{-\frac{3\Phi}{4}} \r d\phi_2 \,,\rc
&&e^8 = \sqrt{g_s N_c \a'} \frac{e^{-\frac{3\Phi}{4}}}{\sqrt{z}} d\s\,, \quad
e^9 = \sqrt{g_s N_c \a'} \frac{e^{-\frac{3\Phi}{4}}}{\sqrt{z}} \s
(d\phi_1 + \cos \tt d\tvf)
\eear
The Killing spinors are
given by:
\be
\e = e^{\frac{\Phi}{8}}e^{-\frac12 \phi_1 \G_{45}+\frac12 \phi_2 \G_{67}}
\eta
\label{Killings}
\ee
where $\eta$ is a constant spinor satisfying:
\be
\s_1 \G_{4567} \eta = -\eta \,,\qquad
\G_{4589} \eta = \eta
\label{Killing}
\ee
The first order equations read:
\bear
g&=& - \r z' \,,
\label{eqforg}\\
e^{2\Phi}&=& \frac{\s}{z \dot z} \,,\label{eqforphi}\\
g' &=& -2 e^{-2\Phi} \r\s \dot \Phi \,,\\
\dot g &=& - z^{-2} e^{-2\Phi} \s g + 2 z^{-1} \r\s e^{-2\Phi} \Phi'\,\,.
\label{eq4}
\eear
It is possible to check that these equations ensure the equation of
motion for the 3-form
$d(e^\Phi\,*F_{(3)}) = 0$.
Notice, however, that these equations are not independent since
(\ref{eq4}) is automatically implied by (\ref{eqforg}) and (\ref{eqforphi}).
The system (\ref{eqforg})-(\ref{eq4}) can be rephrased as a single 
second order 
non-linear partial differential equation (PDE)
for the function $z(\rho,\sigma)$:
\beq
\r z (\dot z - \s \ddot z)=\s (\r \dot z^2 + z' + \r z'')
\label{eqinz}
\eeq
and once $z$ is known, $g$ and $\Phi$ can be obtained from (\ref{eqforg}), (\ref{eqforphi}).
The implicit relation deduced from (\ref{varchange}):
\beq
\s^2 - z e^{-2x} (e^{2z}-\rho^2)=0
\label{dlmimplicit}
\eeq
(where $e^{-2x}$ was defined in (\ref{defs})) solves (\ref{eqinz})
and yields the known solution\footnote{Simple solutions of 
(\ref{eqinz}) are $z=c_1 + c_2 \log \r$ which is not physical since
from (\ref{eqforphi}) one gets $\Phi=\infty$ and 
$z=c_1 \sqrt{2\s^2 + c_2}$ which leads to constant dilaton and
$g=F_{(3)}=0$ and just Ricci flat $\mathbb{R}_{1,5}\times EH_4$ where
$EH_4$ denotes the
Eguchi-Hanson space.}
 (\ref{metric1})-(\ref{defs}).
The value of $g$ in terms of the old variables can
be read by comparing (\ref{C21}) to (\ref{F3}):
\be
g=-\frac{\sin^2\th}{e^{2x} \cos^2{\th} + \sin^2\th}
\label{grelation}
\ee
It is also useful to define quantities $z_0$, $\rho_0$ which depend on $c$ as:
\be
e^{-2x(z_0)} = 0 \,\,,\qquad \rho_0 = e^{z_0}
\label{r0crelation}
\ee
The minimum value that $z$ can attain is $z_0$.
When $c=-1$, then $z_0=0$.

Using (\ref{metric2}) and (\ref{Killing}), it is immediate to see that
 two kinds of D5-brane probes can be
added to the setup preserving the full
supersymmetry of the background. They correspond to the 
$\kappa$-symmetry projectors
$\s_1 \G_{012345}$ and $\s_1 \G_{012389}$. The first kind corresponds
to D5 extended along $x_0,\dots,x_3,\tt,\tvf$ placed at $\s=0$ (notice
that the dilaton does not diverge at $\s=0$ despite the appearance
of (\ref{eqforphi}) since $\dot z\sim \s$ near $\s=0$ as can
be deduced from (\ref{eqinz})). 
These branes, which wrap a compact $S^2$,
are the color branes. Computing the Born-Infeld action on their worldvolume
leads, upon reduction to four dimensions, to the
${\cal N}=2$ SYM action.
In  \cite{DiVecchia:2002ks}, the identification of the Yang-Mills coupling, the $\th_{YM}$
and the radius energy relation was made precise. In order to follow their argument,
notice that at $\s=0$:
\beq
(\s=0, \ \rho<\rho_0) \Rightarrow 
\begin{cases} z=z_0 \cr g = 0 \cr
\end{cases}\,,
\qquad
(\s=0, \ \rho>\rho_0) \Rightarrow 
\begin{cases} z=\log \r \cr g = -1 \cr
\end{cases}\,,
\label{sigma0}
\eeq
where we have used
 (\ref{dlmimplicit}),  (\ref{eqforg}) in order to derive these expressions.
$\s=0, \ \rho\leq\rho_0$ is  
the singular locus of the geometry. As acknowledged in \cite{Gauntlett:2001ps},
it is remarkable that one obtains regular results for physical quantities
despite the singularity.
 
Now,
consider the abelian worldvolume DBI+WZ action of a D5-brane probe
placed at $\s = 0$. Then, expand it to leading order and promote it
to a non-abelian one. Normalizing the $SU(N_c)$ generators as
$tr(T^A T^B) =\frac12 \d^{AB}$, one obtains \cite{DiVecchia:2002ks}:
\beq
S_{YM}=
-\frac{1}{g_{YM}^2}\int d^4x \{ \frac14 F_{\a\b}^A F^{\a\b}_A
+\frac12 D_\a \bar \Psi^A D^\a \Psi_A \} +
\frac{\theta_{YM}}{32\pi^2} \int d^4x  F_{\a\b}^A 
\tilde F^{\a\b}_A
\eeq
where we have defined $\Psi$, which is the complex scalar of the
${\cal N}=2$ vector multiplet as:
\beq
\Psi = \frac{\sqrt{g_s N_c \a'}}{2\pi \a'} \r e^{i \phi_2}
\label{Psidef}
\eeq
After integrating the $S^2$ parameterized
by $\tt,\tvf$, the value of the couplings can be read from the solution
(\ref{sigma0}). For $\rho<\rho_0$ they
are constant while for $\rho>\rho_0$ they read:
\bear
\frac{1}{g_{YM}^2} = \frac{N_c}{4\pi^2} (z|_{\s=0}) =
 \frac{N_c}{4\pi^2} \log \r \,,\label{gym}\\
 \th_{YM} = 2 (g|_{\s=0}) N_c \phi_2 = -2 N_c \phi_2
 \label{tym}
\eear
Now, noticing that the complex scalar $\Psi$ has protected
mass dimension one, one is lead, from (\ref{Psidef}) to the
following radius-energy relation:
\beq
\r = \frac{\m}{\Lambda}
\label{radius-energy}
\eeq
where $\m$ is the mass scale at which the theory is defined and
$\Lambda$ the dynamically generated mass scale.
Then:
\beq
\b(g_{YM})\equiv  \frac{\partial g_{YM}}{\partial \log(\m/\Lambda)}=
\frac{\partial g_{YM}}{\partial \log \r}=
-\frac{N_c}{8\pi^2}g_{YM}^3
\eeq
On the other hand, one can compute the chiral anomaly, which is 
associated to $U(1)_R$ shifts:
\beq
\phi_2 \to \phi_2 + 2\e  \quad \Rightarrow  \quad
\th_{YM} \to \th _{YM} - 4 N_c \e\,\ ,  \qquad \e \in[0,2\pi)
\eeq
The parameter $\e$ takes values in $[0,2\pi)$ since, although in the
geometry one comes back to the same point after
a rotation of $2\pi$ in $\phi_2$, the complete $U(1)_R$ rotation is
of a $4\pi$ angle. This can be seen from (\ref{Killings}), since
the spinors pick a minus sign upon $\phi_2 \to \phi_2 + 2\pi$
\cite{Gauntlett:2001ps}. This is also consistent with the complex 
scalar $\Psi$ having R-charge 2, see (\ref{Psidef}).

Therefore, allowing $\th_{YM}$ to change by a multiple of $2\pi$ one
finds the values of $\e$:
\beq
\e = \frac{2\pi}{4 N_c}k \,, \qquad k=0,1,\dots, 4N_c-1
\eeq
describing the anomalous breaking $U(1)_R \to \mathbb{Z}_{4N_c}$.

Let us finish the section by defining the holomorphic coupling
$\tau=i\frac{4\pi}{g_{YM}^2}+\frac{\th_{YM}}{2\pi}$, which, in view
of (\ref{sigma0}), (\ref{gym}),(\ref{tym}), and defining $u=\Lambda\rho e^{i\phi_2}$,
$u_0=\Lambda\rho_0$  is:
\bear
\tau &=&  i \frac{N_c}{\pi} \log\frac{u_0}{\Lambda} \,,\qquad {\rm for} \ |u|\leq u_0\,,\rc
\tau &=& i \frac{N_c}{\pi} \log\frac{u}{\Lambda} \,,\qquad {\rm for} \ |u|\geq u_0\,
\eear
which is precisely the result expected for a ring-like distribution of vevs
\cite{Gauntlett:2001ps}.

\section{The dual with flavor}
\setcounter{equation}{0}

\subsection{Flavor branes in the solution}
\label{flavorprobes}

The second kind of supersymmetric embeddings for D5-branes is
associated to the $\k$-symmetry projector $\s_1 \G_{012389}$.
It corresponds to branes extended along $x_0,x_1,x_2,x_3,\s,\phi_1$
at any point in the other coordinates $\tt,\tvf,\r,\phi_2$.
These branes provide
fundamental hypermultiplets in order to
build 
${\cal N}=2$ SQCD. 
First of all, since they are extended in the $\s$ direction, their volume
is infinite (compared to the color branes), making exactly zero the
effective four-dimensional gauge coupling living on them,
and therefore providing a global symmetry group $U(N_f)$ if they are placed on top of each other. 
Remember that this (and not $U(N_f)\times U(N_f)$) is the flavor symmetry
of the ${\cal N}=2$ SQCD lagrangian, due to the coupling (in 
${\cal N}=1$ notation)
\beq
Tr(\tilde Q \Psi Q)
\label{coupling}
\eeq
where the $Q, \tilde Q$ represent the (anti)-fundamental chiral multiplets
inside the ${\cal N}=2$ hypermultiplet
and $\Psi$ is the chiral multiplet in the adjoint inside the 
${\cal N}=2$ vector multiplet. This is consistent with
having only flavor D5-branes (anti D5-branes would break supersymmetry).
It can be seen that the coupling (\ref{coupling}) appears on the
worldvolume theory of a probe color D5-brane. The easiest argument is
that ${\cal N}=2$ is preserved, so (\ref{coupling}) is needed. Also, from the
point of view of the brane intersection, it is what one expects, since
$\Psi$ parameterizes the two directions which are orthogonal to 
both color and flavor branes (those spanned by $\rho,\phi_2$).
The values of $\r$, $\phi_2$ at which each flavor brane is placed give the
complex mass of the corresponding hypermultiplet (taking into account
the identifications of section \ref{sect: solution}). The position
in the $\tt,\tvf$ coordinates does not play any role in the low energy
field theory, since we are interested in energies below the inverse radius of the
$S^2$. However, this statement should be taken with certain criticism, since, as
usually happens in the wrapped brane models, one cannot really achieve a separation
of scales between the unwanted Kaluza-Klein modes and the relevant physical modes.
A discussion on this point and some ideas of how to deal with this problem can
be found in \cite{Gursoy:2005cn}.

Let us now focus on the $U(1)_R$ symmetry that, as described in 
section \ref{review},
corresponds to rotations of the $\phi_2$ angle.
Massless flavors would correspond to
branes located at the origin of the moduli space $\rho=0$, a point
which is invariant under $\phi_2$ rotations, and therefore preserves the $U(1)_R$.
A brane located at some $\rho > 0$ would correspond to a 
massive flavor, and, as expected, breaks explicitly the $U(1)_R$. 

There is an $SO(3)$ isometry which acts on $\tt,\tvf,\phi_1$ (it is not $SU(2)$ since
$\phi_1$ takes values between $0$ and $2\pi$ whereas in the usual $SU(2)$ left
invariant one-forms, it would range up to $4\pi$), which does not play
a role in the low energy theory \cite{Gauntlett:2001ps}.
Thus, the $SU(2)_R$ symmetry of the field theory cannot be entirely realized in 
the geometry. Nevertheless, there is another isometry in the solution, the
rotations in the $\phi_1$ angle, which can be identified with the
$U(1)_J \subset SU(2)_R$. An indication that this is the case comes from
comparison to a non-critical string setup. It was argued in \cite{Bigazzi:2005md}
(see also \cite{Israel:2005fn}) that, if one considers the eight-dimensional
background $\mathbb{R}_{1,3}\times \mathbb{R}_2 \times \frac{SL(2,\mathbb{R})_4}{U(1)}$
and places D3-branes extended along the $\mathbb{R}_{1,3}$, at the tip of the cigar,
and D5-branes extended along the $\mathbb{R}_{1,3}\times \frac{SL(2,\mathbb{R})_4}{U(1)}$,
then the low energy 
theory living on the D3 is ${\cal N}=2$ SQCD. Heuristically, one can think of
this non-critical setup as the one described in this paper where the $S^2$
parameterized
by $\tt,\tvf$ has shrunk to string scale size. Naturally, the curvature would be of the order of the
string scale.
In \cite{Murthy:2003es}, it was shown that, by considering the fermions, this $U(1)$
forms part of an $SU(2)$ symmetry. The wrapping of the flavor D5 branes does not break
this symmetry, regardless the hypermultiplet is massive or not. However,
the scalars inside the hypermultiplets are charged under the $SU(2)_R$, so the
meson-like excitations lie in non-trivial representations of this group, as will be shown in 
section \ref{sect:mesons}.

When two hypermultiplets have the same mass, it is possible to enter
a Higgs branch by
giving
expectation values to $Q$, $\tilde Q$. From the gravity side, this should be
described by using the non-abelian brane worldvolume action along the lines
of \cite{Erdmenger:2005bj}.

Table \ref{branescheme} summarizes the setup.

\begin{table}[ht!]
\begin{center}
\begin{tabular}{|c|c|c|c|c|c|c|c|}
\multicolumn{1}{c}{ }
&
\multicolumn{1}{c}{ }
&
\multicolumn{4}{c}{$\overbrace{\phantom{\qquad\ \qquad\quad\qquad\qquad\qquad\qquad
\qquad
\ \ \,}}^{CY_2}$}
&
\multicolumn{2}{c}{$\overbrace{\phantom{\ \quad\qquad\qquad\qquad
\ \ \,}}^{\mathbb{R}_2}$}
\\
\hline
\multicolumn{1}{|c|}{ }
&
\multicolumn{1}{|c|}{$x_{1,3}$}
&\multicolumn{1}{|c|}{$\sigma$}
&\multicolumn{1}{|c|}{$\phi_1$}
&\multicolumn{1}{|c|}{$\tt$}
&\multicolumn{1}{|c|}{$\tvf$}
&\multicolumn{1}{|c|}{$\r$}
&\multicolumn{1}{|c|}{$\phi_2$}
\\
\hline
$N_c$ D5 &$-$&$\cdot$&$\cdot$&$\bigcirc$&$\bigcirc$&$\cdot$&$\cdot$\\
\hline
$N_f$ D5 &$-$&$-$&$\bigcirc$&$\cdot$&$\cdot$&$\cdot$&$\cdot$\\
\hline
\multicolumn{1}{|c|}{ }
&
\multicolumn{1}{|c|}{4D spacetime}
&\multicolumn{1}{|c|}{ }
&\multicolumn{1}{|c|}{$U(1)_J\subset SU(2)_R$}
&\multicolumn{1}{|c|}{ }
&\multicolumn{1}{|c|}{ }
&\multicolumn{1}{|c|}{Energy scale}
&\multicolumn{1}{|c|}{$U(1)_R$}
\\
\hline
{ } &$(-\infty,\infty)$&$[0,\infty)$&$[0,2\pi)$&$[0,\pi]$&$[0,2\pi)$&$[0,\infty)$&$[0,2\pi)$\\
\hline
\end{tabular}
\caption{A scheme of the setup:
for the brane configuration,
a line $-$ means that the brane spans a non-compact dimension, a point
$\cdot$ that it is point-like in that direction and a circle
$\bigcirc$ that it wraps a compact cycle.
The physical meaning of the different dimensions is summarized as
well as their ranges. Above, it is shown which directions spanned the Calabi-Yau
and which the transverse plane before backreaction.
\label{branescheme}}
\end{center}
\end{table}

\subsection{Computing the flavor brane backreaction}
\label{sect:back}

Now that the embedding for flavor branes  has been identified, 
the procedure of \cite{Casero:2006pt} can be followed
in order to find a backreacted solution and the effects of 
having $N_f \sim N_c$ in the field theory.
The unquenched solution is a solution of supergravity coupled to the flavor branes which,
unlike the color branes, do not disappear into fluxes. The global flavor symmetry of the
field theory corresponds to the gauge symmetry on the brane worldvolume. As explained
above, this symmetry becomes global, from the four-dimensional point of view because
the flavor branes have infinite volume since they are extended in the non-compact $\sigma$ 
direction.

\begin{figure}[!htb] 
\begin{center}
   \includegraphics[width=0.45\textwidth]{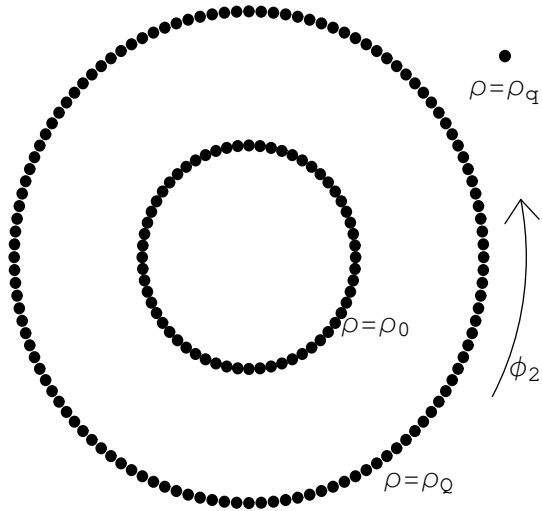} 
\end{center}
   \caption{A scheme of the configuration on the $\rho,\phi_2$ plane, which is the
complex $u$-plane up to a factor of $\Lambda$. 
The inner circle at radius $\rho_0$ is the distribution of $N_c$ vevs for the adjoint scalars.
The outer circle represents the $N_f$ masses of the fundamental hypermultiplets. The associated 
flavor branes are also extended along the transverse $\s,\phi_1$ directions. These smeared
distributions restore the $U(1)_R$ broken by each single vev or mass. The point at a scale
$\r_q$ represents a flavor probe brane whose dynamics will be studied in section
\ref{sect:mesons}. }
   \label{picture}
\end{figure}

Clearly, the solution must depend on the masses one chooses for the fundamental hypermultiplets
(the position in $\rho, \phi_2$ of the branes).
The case addressed below is the simplest one: each single mass 
breaks the $U(1)_R$ symmetry
but since $N_f\to \infty$, we may consider a setup in which the 
masses are smeared and 
distributed in a ring, such that the $U(1)_R$ is restored.
The flavor symmetry is explicitly broken to $U(1)^{N_f}$.
Notice the similarity with the also ring-like distribution of vevs.  
In particular,
an infinitely thin ring at $\rho=\rho_Q$ 
 will be considered, although the construction can be easily generalized
to other configurations with the same symmetry.
The choice of masses fixes the distribution of branes along the $\rho, \phi_2$ directions.
We still have to fix their distribution in $\tt,\tvf$. Since in the low energy theory these coordinates
play no role, one can just consider the simplest possibility, {\it i.e.}, a uniform distribution
such that the $SO(3)$ symmetry is also kept.
This dramatically simplifies the task of writing 
an ansatz. 
 Figure
\ref{picture} depicts the situation.

The D5 flavor branes are magnetic sources for the $F_{(3)}$ RR-field strength.
As in \cite{Casero:2006pt}, a supersymmetric solution can be found by modifying the Bianchi
identity for the $F_{(3)}$ and inserting the suitable ansatz in the supersymmetry variation
equations (\ref{susytransf}).
The worldvolume action for the flavor branes is:
\beq
S_{flavor}=T_5 \sum^{N_f}
\left( -\int d^6x e^\frac{\Phi}{2} \sqrt{-\hat g_{(6)}}
+ \int P[C_{(6)}]\,\,.
\right)
\eeq
We are interested in the WZ part, which, being an infinite
sum can be promoted to a
ten-dimensional integral.
\beq
T_5 \sum^{N_f}\int P[C_{(6)}] \to 
T_5 \int Vol({\cal Y}_4) \wedge C_{(6)}
\label{WZterm}
\eeq
where $Vol({\cal Y}_4)$ is the brane density, subject to the
normalization condition $\int Vol({\cal Y}_4)=N_f$.
For the distribution described above:
\beq
Vol({\cal Y}_4)=\frac{N_f}{8\pi^2} 
 \delta(\rho-\rho_Q)\ \sin \tt d\r \wedge d\phi_2 
\wedge d\tt \wedge d\tvf 
\label{y4}
\eeq
The modified Bianchi identity reads\footnote{In general, if for a form $F_{(n)}=dA_{(n-1)}$
there is an action $-\frac{1}{2n!}\int \sqrt{|g|} F^2
+\int G \wedge A$, the equation of motion for the form reads
$d*F = sign(g)(-1)^{D-n+1} G$. In this case, the relevant part of
the action (go to string frame for this computation) reads:
$-\frac{1}{2\kappa_{(10)}^2}\frac{1}{2\cdot 7!}\int \sqrt{|g|} F_{(7)}^2
+T_5 \int Vol({\cal Y}_4)\wedge C_{(6)}$
so the equation of motion is 
$\frac{1}{2\kappa_{(10)}^2} d*F_{(7)}=- T_5
Vol({\cal Y}_4)$. Taking into account $F_{(3)}=- *F_{(7)}$
and   we arrive at (\ref{newdF}).}:
\beq
dF_{(3)}= 2\kappa_{(10)}^2 T_5 Vol({\cal Y}_4)=
 g_s \a' \ \frac{N_f}{2} 
 \delta(\rho-\rho_Q)\ \sin \tt d\r \wedge d\phi_2 
\wedge d\tt \wedge d\tvf 
\label{newdF}
\eeq
where (\ref{t5k10}) and (\ref{y4}) were used to obtain the second equality.
The natural ansatz satisfying this condition that generalizes (\ref{F3}) is:
\bear
F_{(3)}&=& N_c g_s \a' \left[ -g' d\phi_2 \wedge d\r \wedge
(d\phi_1 + \cos \tt d\tvf) - \dot g d\phi_2 \wedge d\s \wedge
(d\phi_1 + \cos \tt d\tvf)+\right. \rc
&& \left. + (g + \frac{N_f}{2N_c}\Theta(\rho-\rho_Q))
 \sin \tt d\phi_2\wedge d\tt \wedge 
d\tvf \right]\,.
\label{newF3}
\eear
where $\Theta$ is the Heaviside step function.
Maintaining the same ansatz for the metric\footnote{
If one considers a more general form for the metric
\bear
ds_{10}^2 &=& g_s N_c \a' e^\frac{\Phi}{2} \left[ 
\frac{1}{g_s N_c \a'} dx_{1,3}^2 +  z (d\tt^2 +
\sin^2 \tt d\tvf^2) + \right.\rc
&&
\left.
+ \a^2 e^{-2\Phi} (d\rho^2 + \rho^2 d\phi_2^2)
+ \b^2 \frac{e^{-2\Phi}}{z} \left( d\s^2 + \s^2
(d\phi_1 + \cos \tt d\tvf)^2 
\right) \right]\,,\nonumber
\eear
where $\a(\rho,\s),\b(\rho,\s)$, the Killing spinor
equations impose that both $\a$ and $\b$ are constant.
} (\ref{metric2}) and the expressions for the Killing spinors 
(\ref{Killings}),(\ref{Killing}), one finds a
very slight modification of (\ref{eqforg})-(\ref{eqinz}), namely:
\bear
g + \frac{N_f}{2N_c}\Theta(\rho-\rho_Q) &=& - \r z' \,,
\label{neweqforg}\\
e^{2\Phi}&=& \frac{\s}{z \dot z} \,,\label{neweqforphi}\\
g' &=& -2 e^{-2\Phi} \r\s \dot \Phi \,, \label{newgpri}
\\
\dot g &=& - z^{-2} e^{-2\Phi} \s (g+\frac{N_f}{2N_c}\Theta(\rho-\rho_Q)) +
 2 z^{-1} \r\s e^{-2\Phi} \Phi'\,\,.
\label{neweq4}
\eear
As in the unflavored case, (\ref{neweq4}) is implied by (\ref{neweqforg}), (\ref{neweqforphi})
and
it is possible to check that these equations ensure the equation of
motion for the 3-form
$d(e^\Phi\ {}^*F_{(3)}) = 0$. The generalization of (\ref{eqinz}) is
\beq
\s \frac{N_f}{2N_c}\delta(\rho-\rho_Q) +
\r z (\dot z - \s \ddot z)=\s (\r \dot z^2 + z' + \r z'')\,\,.
\label{neweqinz}
\eeq
This equation seems very difficult, if possible at all, 
 to solve in general. However, the important point is that it
uniquely defines the relevant solution. For $\rho < \rho_Q$, the solution should be equal
to the unflavored one, which is explicitly known (\ref{dlmimplicit}). A first argument to
see this comes from an electrostatic analogy: a spherically symmetric 
distribution
of charges does not alter the field in its interior. 
Such a reasoning was used
in \cite{Kiritsis:2002xr} in order to find the supergravity solution 
corresponding to shells
of five-branes uniformly distributed  on $S^3$ spheres. 
Notice the 
similarity between such a
setup and the one discussed in this note. A second 
argument comes from field theory and
holomorphic 
decoupling (see the final part of section
\ref{sect:features}).

\begin{figure}[!htb] 
   \includegraphics[width=0.45\textwidth]{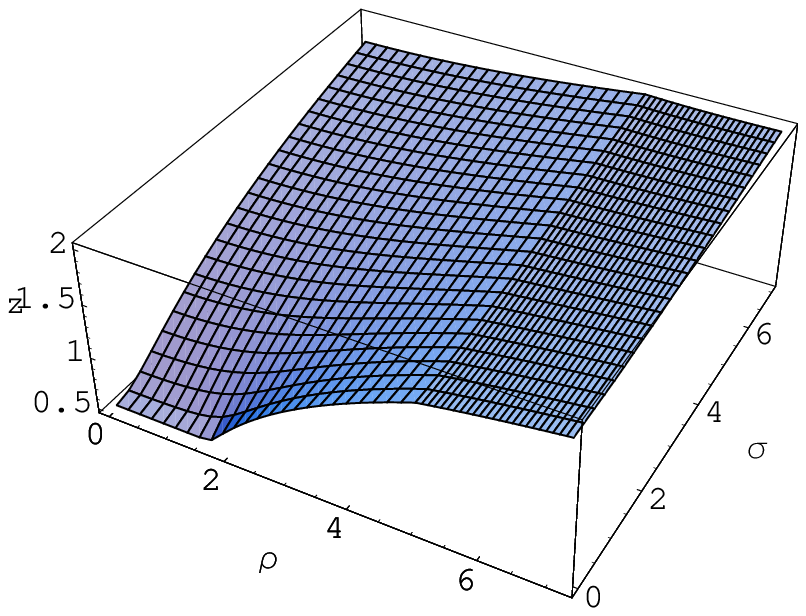} 
   \includegraphics[width=0.45\textwidth]{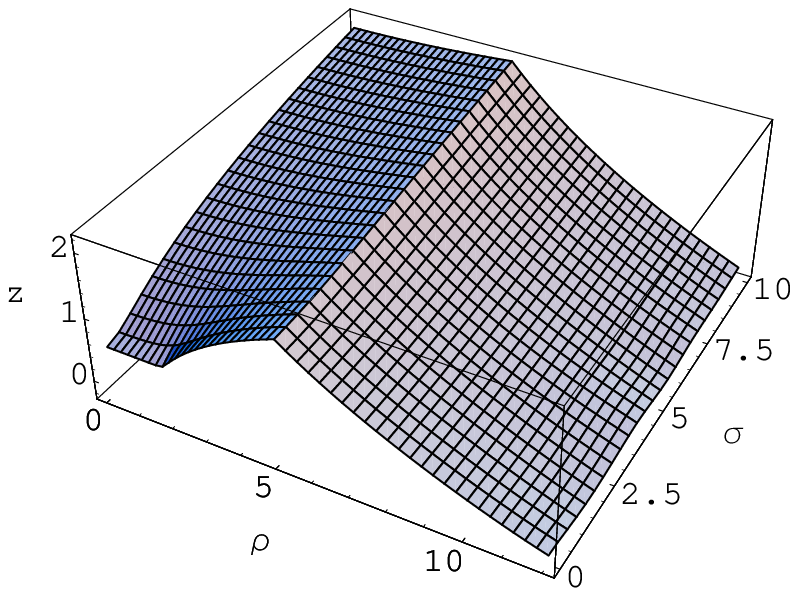} 
   \caption{Numerical approximations to the function $z(\r,\s)$
   that enters the metric. Both plotted solutions correspond to
   fixing $c=0$ ($\rho_0=\sqrt{e}$) and $\r_Q=5$. It is apparent
   from the graphics the qualitative change of behavior at such
   values of $\r$. On the left $N_f = N_c$. On the right, $N_f = 6 N_c$,
   a non-asymptotically free field theory that 
   eventually hits a Landau pole. 
   This corresponds to $z$
   becoming zero at some value of $\r$, as shown in the graph.}
   \label{numz}
\end{figure}

Thus, we are left with the problem of finding the function $z(\r,\s)$ for $\rho>\rho_Q$.
It will be shown in section \ref{sect:features} that $z(\r,0)$ can be written in a simple form
and leads to several gauge theory predictions. By continuity, and regarding the
previous paragraph $z(\r_Q,\s)$ must be the same as in the unflavored case and therefore is
known. From (\ref{neweqforg}), one can read how the derivative changes when 
traversing the brane shell, fixing $z'(\r_Q,\s)$. This set of boundary conditions fixes the
solution. In figure \ref{numz}, a numerical approximation to the function
$z(\r,\s)$ in two particular
cases is presented.

There is a last subtle point that requires discussion
before closing this section. 
We have introduced a number of fundamentals and found new equations that define a new solution.
If this new solution still describes a theory with $N_c$ colors, equation
(\ref{quant}) should still hold.
Let us now prove so.
Comparing to (\ref{quant2}), the integrals in $\phi_1,\phi_2$ are trivial, but the angle
$\theta$ is not manifest in the geometry any more. This problem is solved by noticing that
the components of $F_{(3)}$ relevant to this integration (those not containing 
$d\tt$ or $d\tvf$ can be written as (see (\ref{newF3})) $N_c g_s \a' dg\wedge d\phi_2\wedge d\phi_1$
and that, at infinity, the limits of integration in $\theta$ correspond to $(\s=0,\r=\infty )$ and $(\s=\infty ,\r=0 )$.
Then:
\be
\int F_{(3)} = 4\pi^2 g_s N_c \a' ( g|_{((\s=\infty ,\r=0 )} - g|_{(\s=0,\r=\infty )})=4\pi^2 g_s N_c \a'
\label{quant3}
\ee
where the equalities $g|_{(\s=\infty ,\r=0 )}=0$ and $g|_{(\s=0,\r=\infty )}=-1$ have been used, and also
the fact that $g$ is continuous along the integration path. The first 
equality is trivial since for $\rho<\rho_Q$ the solution coincides with the unflavored one so 
$g$ can be read from (\ref{grelation}) setting $\theta =0$. The second one will 
be proved in section \ref{sect:features}. Thus, since (\ref{quant3}) coincides with (\ref{quant2}),
the quantization condition still holds in the flavored case.

\subsection{Gauge theory features}
\label{sect:features}

Let us now study the gauge theory implications of this computation.
The $\rho<\rho_0$ region is unchanged with respect to section \ref{review}
when the fundamentals are present,
so in the following, only the $\rho>\rho_0$ region is considered.
By repeating the argument of \cite{DiVecchia:2002ks}
reviewed in section \ref{sect: solution}, it is straightforward
to generalize (\ref{gym}), (\ref{tym}) to:
\bear
\frac{1}{g_{YM}^2} &=& \frac{N_c}{4\pi^2} (z|_{\s=0})\,,\label{newgym}\\
 \th_{YM} &=& 2 \Big(g|_{\s=0} + \frac{N_f}{2N_c}\Theta(\rho-\rho_Q)\Big) N_c \phi_2 
 \label{newtym}
\eear
 Equation (\ref{newgpri}) implies
 that $g|_{\s=0}$ is a constant. Repeating the argument that
the solution at small $\r$ should be the same as the unflavored one,
we find (as in (\ref{sigma0})) $g|_{\s=0} = -1$. Thus:
\beq
\th_{YM}  = -2N_c \Big(1 - \frac{N_f}{2N_c}\Theta(\rho-\rho_Q)\Big) \phi_2
\label{flavtym}
\eeq
One can compute the monodromy of the $\th_{YM}$ angle when making a full
$U(1)_R$ rotation through the space of vevs (at a fixed modulus $\rho \Lambda$):
\beq
\phi_2 \to \phi_2 + 4\pi  \quad \Rightarrow  \quad
\frac{\th_{YM}}{2\pi} \to \frac{\th _{YM}}{2\pi} - 2 \Big(2 N_c-N_f \Theta(\rho-\rho_Q)\Big) 
\eeq
which depends on $N_f \Theta(\rho-\rho_Q)$, 
the number of fundamental hypermultiplet masses encircled by the path.
This result is in exact agreement with the field theory expectations 
(for a review, see \cite{Peskin:1997qi}).

Let us now turn to $g_{YM}$. Using $g|_{\s=0} = -1$,
it is immediate to integrate (\ref{newgym}).
Imposing
that $z$ matches the unflavored 
solution
for $\r < \r_Q$ and is continuous at $\rho=\r_Q$, one obtains:
\be
\frac{1}{g_{YM}^2}=\frac{1}{4\pi^2}\Big[
\Big(N_c - \frac{N_f}{2} \Theta(\r-\r_Q)\Big) \log \r
+ \frac{N_f}{2} \Theta(\r-\r_Q) \log \r_Q\Big]
\label{flavgym}
\ee
Using (\ref{radius-energy}), the $\beta$-function at an
energy scale $\m$ is straightforwardly computed: $
\b (g_{YM})(\m) = -\frac{g_{YM}^3}{8\pi^2} (N_c - \frac{N_f(\m)}{2})$
where $N_f (\m)$ is defined as the number of flavors for which the
modulus of their masses is
smaller than the scale. The fact that matter fields with bigger mass
do not contribute is usually an approximate statement, but, due to the
radial symmetry, in this case it is exact as will be shown below.
Their effect is just to modify the dynamically generated scale
$\Lambda$.
For $N_f \geq 2N_c$, the theory loses asymptotic freedom and
for $N_f > 2N_c$, it eventually hits a Landau pole. In the gravity solution,
this is translated in the fact that when $N_f > 2N_c$, the function
$z$ vanishes at some finite $\rho$ and the metric becomes singular.
An analogous behavior in a D3D7 solution was discussed in \cite{Kirsch:2005uy}.

The holomorphic coupling can be written by using (\ref{flavtym}) and (\ref{flavgym})
(define $u_Q=\r_Q \Lambda$):
\bear
\tau &=&  i \frac{N_c}{\pi} \log\frac{u_0}{\Lambda} \,,\qquad {\rm for} \ |u|\leq u_0\,,\rc
\tau &=& i \frac{N_c}{\pi} \log\frac{u}{\Lambda} \,,\qquad {\rm for} \ u_0\leq |u|\leq u_Q\,,\rc
\tau &=& i \frac{N_f}{2\pi} \log\frac{u_Q}{\Lambda}
+\frac{i}{\pi}(N_c - \frac{N_f}{2}) \log\frac{u}{\Lambda} \,,\qquad {\rm for} \ |u|\geq u_Q\,,
\label{flavoredtau}
\eear
This behavior precisely matches the field theory result. 
The proof is a simple
generalization of the one of \cite{Gauntlett:2001ps} for the unflavored case.
The coupling is given in terms of the so-called prepotential.
As argued in \cite{Buchel:2000cn}, non-perturbative 
contributions to the prepotential
vanish in the large $N$ limit as long as one probes the 
theory far enough from the
vevs $|u - a_i| \gg N_c^{-1}$ (or, in this case, 
also from the 
masses of the hypermultiplets).
The perturbative prepotential can be read, for instance, from 
\cite{Marshakov:1997cr}. Slightly adapting
conventions in order to match those of \cite{Gauntlett:2001ps}:
\be
{\cal F}= \frac{i}{8\pi}\sum_{i\neq j} (a_i - a_j)^2 \log \frac{(a_i -a_j)^2}{\Lambda^2}
- \frac{i}{8\pi}\sum_{\a=1}^{N_f} \sum_{i=1}^{N_c} (a_i - m_\a)^2  \log \frac{(a_i -m_\a)^2}{\Lambda^2}
\ee
The gauge group is broken to $U(1)^{N_c}$ due to the vevs of the scalars. We now probe the theory
by considering all but one of the vevs in  the ring-like distribution and moving around the other one, 
$u$, so we have $U(1)^{N_c-1}\times U(1)$ where the last $U(1)$ corresponds to 
the probe brane. In the large $N_c$ limit, the coupling is:
\be
\tau (u)= \frac{\partial^2 {\cal F}}{\partial u^2} =\frac{i}{2\pi}\sum_{i=1}^{N_c} \log\frac{(u-a_i)^2}{\Lambda^2}
- \frac{i}{4\pi}\sum_{\a=1}^{N_f} \log\frac{(u-m_\a)^2}{\Lambda^2}
\ee
In the large $N_c,N_f$ limit, the sum over discrete 
distributions can be replaced by integrals, namely:
\be
\tau(u)= \frac{i}{2\pi}\int d^2a {\cal \rho}_{<\Psi>} (a) \log \frac{(u-a)^2}{\Lambda^2}
- \frac{i}{4\pi}\int d^2m {\cal \rho}_{m_Q} (m) \log \frac{(u-m)^2}{\Lambda^2}
\label{tauint}
\ee
where ${\cal \rho}_{<\Psi>} (a)$, ${\cal \rho}_{m_Q} (m)$ are 
the density distributions of vevs for the
adjoints and masses for the fundamentals respectively. The 
densities must satisfy the normalization
$\int d^2a  {\cal \rho}_{<\Psi>} (a)= N_c$, $\int d^2m  
{\cal \rho}_{m_Q} (m)= N_f$. 
Therefore, the ring-like distributions are  
${\cal \rho}_{<\Psi>} (a)=\frac{N_c}{2\pi u_0}\delta(|a|-u_0)$ and
${\cal \rho}_{m_Q} (m)=\frac{N_f}{2\pi u_Q}\delta(|m|-u_Q)$. 
Performing the integrals (\ref{tauint}) and adjusting
an additive constant that can be reabsorbed as
a rescaling of $\Lambda$
one finds precisely (\ref{flavoredtau}). The rescaling of $\Lambda$
is what one usually obtains from holomorphic decoupling.

In the context of fractional branes at orbifolds with 
massless fundamental hypermultiplets,
results similar to those reported in this section were 
presented in \cite{Bertolini:2001qa,Bertolini:2002xu}.

\section{Mesonic excitations}
\label{sect:mesons}
\setcounter{equation}{0}

This section is devoted to the analysis of the meson-like excitations,
both in the quenched and unquenched backgrounds.
In order to address this question, let us add an additional flavor 
to the theory
with a mass defined by its position in $\r,\phi_2$. The flavor
group is now $U(1)^{N_f +1}$. Obviously, the effect in the geometry
of this new brane is suppressed by a power $N^{-1}$ and is negligible.
Thus, its excitations can be described by considering it a probe in 
the background described in the previous sections.

The spectrum can be computed by expanding the action to 
quadratic order in the fluctuations.
The analysis is
similar to that in \cite{Kruczenski:2003be}, where
also an ${\cal N}=2$ theory was analyzed. There are, however, important
differences in the low energy field theory: apart from having a large number
of hypermultiplets,
here there is no massless adjoint
hypermultiplet.

Let us consider a probe brane extended along
$x_0,x_1,x_2,x_3,\s,\phi_1$ and sitting at a point defined by
$\tt_q,\tvf_q,\rho_q,{\phi_2}_q$. Its worldvolume action reads
 (below, string frame will be used,
so the metric (\ref{metric2}) should be multiplied by $e^{\frac{\Phi}{2}}$):
\beq
S_{D5}= T_5 \left( -\int d^6 \x e^{-\Phi} \sqrt{-P[g]+ 2\pi \a'   F} 
+ \int P[C_{(6)}] + \frac{(2\pi \a')^2}{2}
 \int P[C_{(2)}]\wedge F \wedge F \right)
\label{probeaction}
\eeq
although the last term does not contribute at quadratic order, 
as can be seen by choosing a gauge in which
$C_{(2)}= g_s N_c \a' 
\left(g+\frac{N_f}{2N_c}\Theta(\r-\r_Q)
\right) d\phi_2 \wedge (d\phi_1 + \cos \tt d\tvf )$.
We  need the expression for the $C_{(6)}$ in the background which
is defined as $dC_{(6)} = F_{(7)} = - *F_{(3)}$. 
Taking the Hodge dual of (\ref{newF3}), one finds:
\bear
F_{(7)} = g_s N_c \a' dx_0\wedge dx_1\wedge dx_2\wedge dx_3\wedge
\Big[-\frac{g'e^{4\Phi}z}{\r\s}\sin\tt d\tt\wedge d\tvf\wedge d\s
+  \rc
 + \frac{\dot g z^2 e^{4\Phi}}{\r\s}\sin\tt d\tt\wedge d\tvf\wedge d\r
 -\big(g+\frac{N_f}{2N_c}\Theta(\r-\r_Q)\big)
\frac{\s}{\r z^2} d\r\wedge d\s \wedge (d\phi_1 + \cos \tt d\tvf )
\Big]
\eear
Using (\ref{neweqforg})-(\ref{neweq4}), one can write the following
simple expression for the RR-potential:
\beq
C_{(6)} = g_s N_c \a' dx_0\wedge dx_1\wedge dx_2\wedge dx_3\wedge
\left( z e^{2\Phi} \sin\tt d\tt \wedge d\tvf
-\frac{\s}{z} d\s\wedge (d\phi_1 + \cos \tt d\tvf ) \right)
\eeq

\subsection{Fluctuation of the scalar fields}
In order to describe the small fluctuations, the simplest method is
to consider the static gauge and use $x_\m , \s, \phi_1$ as worldvolume
coordinates.
The embedding can be written as:
\beq
\tt = \tt_q + \d\tt \,\,, \quad
\tvf = \tvf_q + \d\tvf \,\,, \quad
\r = \r_q + \d\r \,\,, \quad
\phi_2 = {\phi_2}_q + \d\phi_2 \,\,, \quad
\eeq
where all the fluctuations depend on $x_\m , \s, \phi_1$.
It is immediate to
see that at quadratic order, these scalar fluctuations do not mix with those
of the worldvolume gauge field which will be studied in the next section.
Expanding the action (\ref{probeaction}) up to quadratic order 
 one finds:
\bear
S&=& T_5 g_sN_c\a'\int d^4x d\phi_1 d\s \Big[ \frac12 \frac{\s}{z}e^{-2\Phi}
g_s N_c \a' (\dx \d\r)^2 + \frac12 \s (\ders\d\r)^2+\frac{1}{2\s} (\dph\d\r)^2+\rc
&&+\frac12 \r_q^2\frac{\s}{z}e^{-2\Phi}
g_s N_c \a'(\dx \d\phi_2)^2 + \frac12 \r_q^2\s (\ders\d\phi_2)^2+\frac{1}{2\s} \r_q^2
(\dph\d\phi_2)^2+
\rc
&&+\frac12 g_s N_c \a' \s
(\dx \d\tt)^2 + \frac12 \s ze^{2\Phi}(\ders\d\tt)^2+
\frac{ze^{2\Phi}}{2\s} (\dph\d\tt)^2+\rc
&&+\frac12 g_s N_c \a' \s \sin^2 \tt_q
(\dx \d\tvf)^2 + \frac12 \s ze^{2\Phi}\sin^2 \tt_q (\ders\d\tvf)^2+
\frac{ze^{2\Phi}}{2\s} \sin^2 \tt_q (\dph\d\tvf)^2+\rc
&&+ze^{2\Phi} \sin \tt_q \left((\ders \d\tt)(\dph \d\tvf)
 - (\dph \d\tt)(\ders \d\tvf)
\right)\Big]
\label{lfluct}
\eear
where $z,\Phi$ should be understood as  functions of $\s$ at $\r=\r_q$. We have defined
$(\dx \d\r)^2 = \eta^{\m\n}(\partial_\m\d\r)(\partial_\n \d\r)$
where $\m,\n$ span the four Minkowski directions. 
 The fluctuations of
$\d\r,\d\phi_2$ are already decoupled and it is very easy to decouple
those for $\d\tt,\d\tvf$. Let us expand the fields as:
\bear
&&\d\r = \chi_1 (\s) e^{i k \cdot x} \cos (l \phi_1)\,\,,\rc
&&\d\phi_2 = \chi_2 (\s) e^{i k \cdot x} \cos (l \phi_1)\,\,,\rc
&&\d\tt = \frac{\chi_+ (\s)+ \chi_-(\s)}{2} 
\sin \tt_q e^{i k \cdot x} \sin(l\phi_1)
\,\,,\rc
&&\d\tvf =  \frac{\chi_+ (\s)- \chi_-(\s)}{2} 
 e^{i k \cdot x} \cos(l\phi_1)
\,\,.
\label{fldefs}
\eear
The equation of motion for $\chi_1,\chi_2$ reads:
\beq
\frac{\s}{ze^{2\Phi}} \bar M^2 \chi_i + \ders ( \s \ders \chi_i) -
\frac{l^2}{\s}\chi_i =0 \,\,.
\label{chiieq}
\eeq
where we have defined the four-dimensional mass $M^2 = - k_\mu k^\mu$
and $\bar M^2 = g_s N_c \a' M^2$.

The equations for $\chi_\pm$ read:
\beq
\bar M^2 \chi_\pm + \frac{1}{\s}\ders(\s ze^{2\Phi} \ders \chi_\pm)
-\frac{l^2 z e^{2\Phi}}{\s^2} \chi_\pm \mp \frac{l}{\s} \ders( z e^{2\Phi})
\chi_\pm=0
\label{chipmeq}
\eeq

\subsection{Fluctuation of the gauge field}
\label{gaugefluct}

The equation of motion for the gauge field at linear order is:
\beq
\partial_a (e^{-\Phi} \sqrt{-\det(P[g])} F^{ab})=0
\label{gfeq}
\eeq
where $a,b$ run over the six worldvolume coordinates and, again, 
 $\mu,\n$ over the Minkowski four-space and are contracted
with the flat metric $\eta_{\m\n}$.
Let us now choose a Lorentz gauge $\partial_\m A^\m =0$.
The following relation is found:
\beq
\dph A_{\phi_1} = -\s \ders (\s A_\s)
\eeq
Defining a transverse  polarization tensor  $\e_\m$ and expanding:
\bear
A_\m = \e_\m \chi_3 (\s) e^{i k\cdot x} \cos (l\phi_1)\,\,,\rc
A_\s = \frac{1}{\s} \chi_4 (\s) e^{i k\cdot x} \cos (l\phi_1)\,\,.
\eear
one finds from (\ref{gfeq}) that $\chi_3,\chi_4$ also
satisfy equation (\ref{chiieq}).

\subsection{Analysis of the spectrum}

It is interesting to notice certain similarity of the
equations (\ref{chiieq}), (\ref{chipmeq}) with (3.6) and (3.31)
of \cite{Kruczenski:2003be}, respectively. Thus, the
analysis of the spectrum is also similar to the one in  \cite{Kruczenski:2003be}.

However, the $l$ we have used here is not the eigenvalue of some
$J^2$ operator acting on the $SU(2)_R$ generators, but it is the
charge under a $U(1)_J\subset SU(2)_R$, {\it i.e.}, the eigenvalue
of a $J_3$ operator. The relation between the two would require a better understanding.
 The bosonic content of the 
generic ($l\geq 2$) ${\cal N} =2$ massive vector multiplet
is given by a massive vector and three real scalars in the $\frac{l}{2}$
of $SU(2)_R$
and two real scalars in the $\frac{l - 2}{2}$ and $\frac{l+2}{2}$.

Since (\ref{chiieq}) is satisfied by $\chi_i$, $i=1,2,3,4$,
we have in fact a massive vector and three real scalars
with the same mass in the same representation of $SU(2)_R$.
 In order to complete the multiplet,
two more real scalars with the same mass are needed.
They are the $\chi_\pm$ of (\ref{chipmeq}). 
This can be shown by mapping 
equation (\ref{chiieq}) to
(\ref{chipmeq}), following a procedure used in 
\cite{Myers:2006qr}
for a similar case.
In fact:
\bear
\chi_{+,(l=L)} = \s^{-L-1} \ders(\chi_{i,(l=L+1)} \s^{L+1})\rc
\chi_{-,(l=L)} = \s^{L-1} \ders(\chi_{i, (l=L-1)} \s^{1-L})
\label{identification}
\eear
In order to prove this, substitute in (\ref{chiieq}) $l \to L\pm 1$ and
define $F_\pm = \s^{1\pm L}\chi_i$. Multiply the equation by
$ze^{2\Phi} \s^{\pm L}$ and derive with respect to $\s$.
Now, substituting $\ders F_\pm = \chi_\pm \s^{1\pm L}$
and multiplying by $\s^{-1\mp L}$, one arrives at (\ref{chipmeq}) with $l=L$.

Thus, the 
bosonic fluctuations yield the bosonic content of
a tower of ${\cal N}=2$ massive vector multiplets which can be identified
with the mesonic excitations of the theory. 
The fermionic spectrum can be inferred from supersymmetry
or could be directly computed along the lines of \cite{Kirsch:2006he}.

The physical, mesonic, excitations should be those satisfying 
equations (\ref{chiieq}) or (\ref{chipmeq}) and moreover being regular
at $\s=0$ and normalizable as $\s \to \infty$.
Normalizability is a problematic issue in this setup due to 
ill-behaved UV
and will be discussed in
section \ref{sect:estim}.

About regularity at the origin, one would need that the
$\chi$ functions are finite and 
even (odd) around  $\s=0$ when $l$ is 
even (odd). From (\ref{chiieq}), one sees that near $\s=0$,
$\chi_{i,l} = c_1 \s^l + c_2 \s^{-l}$. Clearly, regularity selects $c_2=0$
and $\chi_{i,l} \propto \s^l$ which in fact satisfies the criterion of parity
stated above. For the $\chi_+$ modes, substituting in 
(\ref{identification}), we find  $\chi_{+,l}\propto \s^l$ which  is also
fine with the parity condition. Checking the $\chi_-$ modes, involves an
extra subtlety since substituting the leading $\chi_i$ behavior in
(\ref{identification}) one just gets zero. But from (\ref{chiieq}), noting that
near $\s=0$, $\dot z \propto \s$ we see that both $z$ and $e^{2\Phi}$ behave
as $k_1 + k_2 \s^2$ where $k_1,k_2$ are some constants.
Thus, the subleading behavior of $\chi_i$ is described by
$c_1 \s^l (1 +k_3 \s^2)$, so substituting in
(\ref{identification}) one also finds $\chi_{-,l} \propto \s^l$
which, again, is fine.

\subsection{Estimate of the mass spectrum}
\label{sect:estim}

In D5-brane solutions, like the one we are dealing 
with or the Maldacena-N\'u\~nez
background, the dilaton diverges in the UV.  The
supergravity formalism  cannot be trusted
in that region, and the UV completion
of the field theory is a little string theory. Therefore, it is not a surprise that there
are problems in the UV when trying to determine the spectrum. In fact, 
there is no normalizable mode. However, physically, one would expect to have some
modification of the UV that cures this problem, since in the dual theory there are,
of course, physical excitations. In this section, a regularization procedure
similar to those in \cite{Ametller:2003dj,Nunez:2003cf} is proposed, which
basically consists of cutting off the ill-behaved region. It is curious to notice
the qualitative difference with \cite{Caceres:2005yx}, where 
glueballs in Maldacena-N\'u\~nez
were considered. In that case, even though there was no normalizable mode, 
the authors found that
there existed one leading and one subleading mode in the UV and  
identified the 
subleading as the physical one
(in a different context, the same criterion was also proposed
in \cite{Arean:2005ar}). In the present case, there is no 
leading UV mode because
the two modes behave as sine and cosine of some function of $\s$.

In order to justify the procedure, it is interesting to notice the similarity in the
UV of this setup to the D5D5 brane intersection in flat space. Of course, in
such a case, there are also no normalizable modes 
\cite{Myers:2006qr,Arean:2006pk}.
But in the UV,  by a chain of dualities one can connect the system to an M5M5
intersection which indeed has physical normalizable oscillations \cite{Arean:2006pk}.
Let us define:
\beq
y = \log \s
\eeq
such that equation (\ref{chiieq}) can be written in the Schrodinger form:
\beq
\frac{d^2\chi_i}{dy^2} - V(y) \chi_i = 0\,\,.
\eeq
The potential reads:
\beq
V(y)= l^2 - \bar M^2 \frac{e^{2y}}{z e^{2\Phi}}
\label{pot}
\eeq
where $y \in (-\infty,\infty)$. The function $z e^{2\Phi}$ is implicitly known 
in the quenched case (see section \ref{review}) and can be determined numerically
in the unquenched case, as described in \ref{sect:back}.

\begin{figure}[!htb] 
   \includegraphics[width=0.45\textwidth]{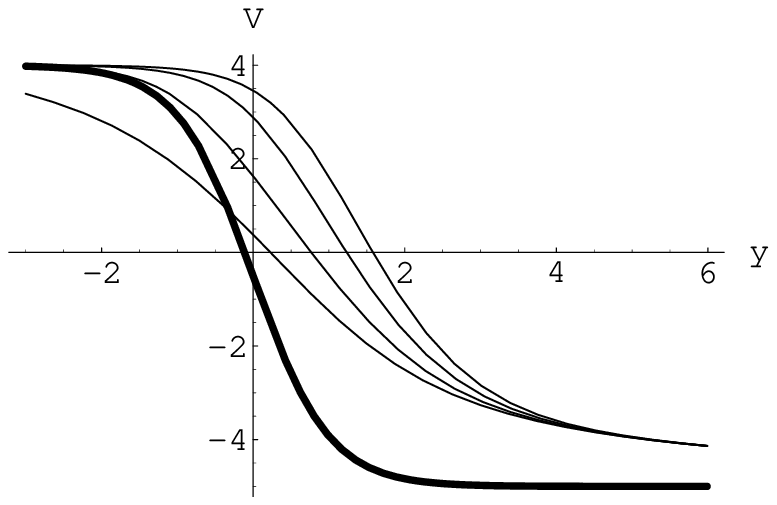} 
   \includegraphics[width=0.45\textwidth]{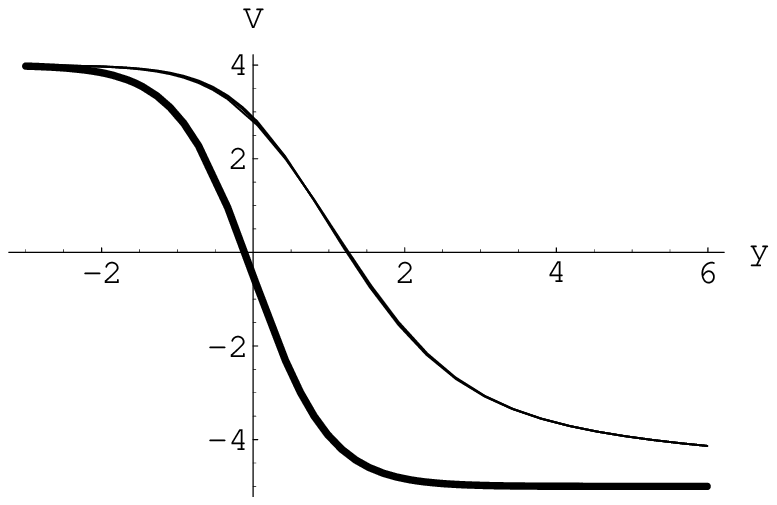} 
   \caption{The figure in the left represents the potential
   (\ref{pot}) with fixed $\mbox{$\frac{N_f}{N_c}=0$},\ \mbox{$l=2$},
   \ \mbox{$\bar M=3$},\ c=-1$ for different values
   of $\r_q=1,2,3,4$ (the biggest the $\r_q$ the upper is the line).
   The thick line corresponds to (\ref{flatpot}).
   The figure on the right is for fixed 
   $\frac{N_f}{N_c}=0,\ \mbox{$l=2$},\ \mbox{$\bar M=3$},\ \r_q=3$ for
   $c=-1,0,1$. The different lines are almost coincident and the
meson masses
   should not strongly depend on $c$.}
   \label{potential}
\end{figure}

For D5D5 intersecting in flat space \cite{Arean:2006pk}
the potential reads:
\beq
V(y) =  l^2 - \bar M^2 \frac{e^{2y}}{e^{2y}+1}
\label{flatpot}
\eeq
Figure \ref{potential} shows the qualitative similarity between the two cases.
For simplicity, the potential (\ref{pot}) is plotted for the quenched case,
although the qualitative behavior for the unquenched one is the same.
From the figure, we see that starting from $y\approx 4 \Rightarrow \s \approx 50$,
all the potentials coincide, {\it i.e.}, the IR effects are no longer important.
It is natural to choose such a scale as a cutoff.
In the plot, the UV problem is apparent since $V(y\to \infty)<0$
and such a Schr\"odinger potential cannot have a discrete spectrum.

 The masses can be estimated  by applying the WKB method, which
 should be reliable for large excitation number.
Let us focus on equation
(\ref{chiieq}). In the notation of appendix \ref{appendix},
we have:
\bear
f=\s &\Rightarrow& f_1=f_2=s_1=r_1=1 \,\,, \rc
h= \frac{\s}{z e^{2\Phi}} &\Rightarrow& s_2=1\,\,, \ \ r_2=-1\,\,,\rc
p = -\frac{l^2}{\s} &\Rightarrow& p_1=p_2 =-l^2\,\,,\ \ s_3=r_3=-1
\eear
where it  was used that $z e^{2\Phi}\sim \s^2$ (up to logarithms) for large $\s$.
In fact, the problem for large $\s$ appears in this formalism
due to the fact that $r_1-r_2-2$ is zero, but it should be a positive
number in order to apply the WKB. Now, suppose
that we can change the UV by that of the M5M5 intersection, which
would have $r_2=-2$ \cite{Arean:2006pk}
and leave all the rest unchanged. With this,
the WKB approximation to the masses reads (\ref{mwkb}):
\beq
\bar M_{WKB}^2 \approx \frac{\pi^2}{\xi^2}n(n-1+3l)\,,\qquad n\geq 1
\eeq
where (\ref{xidef}):
\beq
\xi = \int d\s \sqrt{\frac{h}{f}} = \int_0^{\s_{cutoff}} d\s (z e^{2\Phi})^{-\frac12}
\eeq
Now, the problem for large $\s$ appears in the fact that this integral
is divergent if $\s_{cutoff} \to \infty$. As said above, a natural scale to 
cut the integral is around $y\approx 4\Rightarrow \s \approx 50$.
Although this involves some arbitrariness, everything is fixed in terms of
a single parameter $\s_{cutoff}$ and thus, it should be possible to estimate
how the meson masses depend on the physical parameters $c,\r_q,N_f,\r_Q$.

\begin{figure}[!htb] 
\begin{center}
   \includegraphics[width=0.45\textwidth]{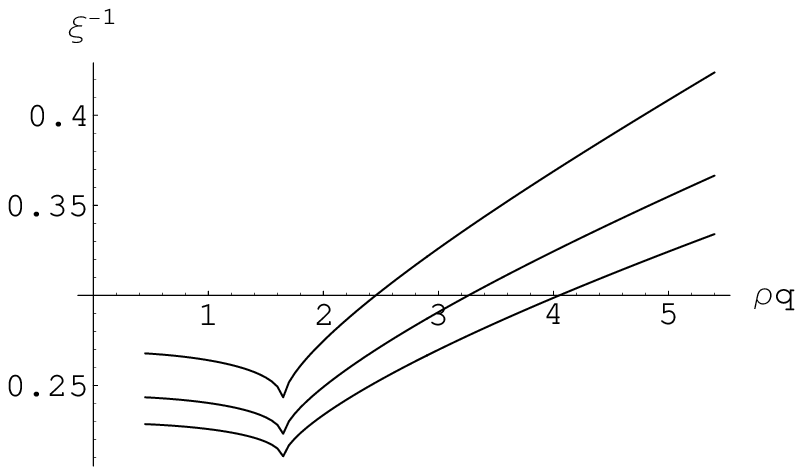} 
   \includegraphics[width=0.45\textwidth]{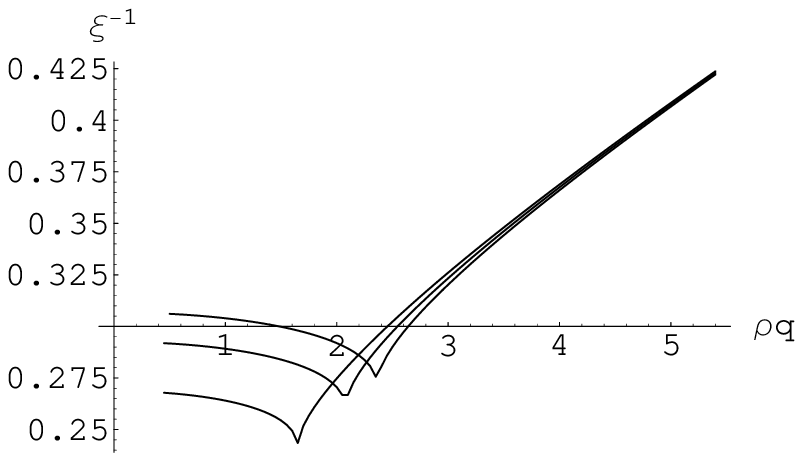} 
\end{center}
   \caption{On the left, 
   $\xi^{-1}$ versus $\r_q$ for $c=0$. Starting from above,
   the three lines correspond to  $\s_{cutoff}=50,70,90$.
   On the right, $\xi^{-1}$ versus $\r_q$ with fixed $\s_{cutoff}=50$
   for $c=0,2,4$. In both plots, $\frac{N_f}{N_c}\to 0$. }
   \label{masses}
\end{figure}

Figure \ref{masses} shows such a dependence
 in the quenched case, which is easier to deal with since
 there is an  expression for $z$ (\ref{dlmimplicit}) and no numerical
 integration of a system of PDEs is needed.
The quantity
$\xi^{-1}$, proportional to the meson mass, is plotted versus $\r_q$.
The graph on the left shows how changing $\s_{cutoff}$
does not change the qualitative properties of the graph, although
of course shifts $\xi$. The graph on the right, shows the
behavior for different values of $c$ (related to $\r_0$ by
(\ref{r0crelation})). For $\r_q \gg \r_0$, the curves become degenerate
as expected on general grounds. The minimal value for the
meson masses is attained
at $\r_q = \r_0$. It is natural to conjecture that this is due to
the fact that when $<\Psi> = m_q$ for some of the eigenvalues,
there are cancellations between
the superpotential term (\ref{coupling}) and the mass term
yielding effectively massless quarks.

An  important question is how the quantum effects
 produced by unquenched flavor
affect the spectrum of physical excitations.
From the discussion of section \ref{sect:back}, it is clear
that for $\r_q \leq \r_Q$ the spectrum is unchanged by the
unquenched fundamentals.
When $\r_q > \r_Q$, the computation
 requires a delicate numerical analysis of the PDE (\ref{neweqinz})
in order to obtain the function $z e^{2\Phi} (\s)$ for given
($\frac{N_f}{N_c},\r_Q,\r_q,c$).
I could not find any significative change of the meson masses 
when the
$N_f \sim N_c$ hypers are introduced, even when $\r_q > \r_Q$.
However, this conclusion
should be taken with a grain of salt due to the difficulty of obtaining a
precise numerical solution  near $\s =0$ and also for $\r_q \gg \r_Q$.

\section*{Acknowledgments}
I am specially indebted to R. Casero and U. G\"ursoy 
for collaboration at certain stages of this project and many useful
and encouraging
discussions. Thanks also to C. N\'u\~nez for a critical reading 
of the manuscript and valuable comments.
This work was supported by European Commission Marie 
Curie Postdoctoral Fellowships, under contract  
MEIF-CT-2005-023373. It was also partially supported by
INTAS grant 03-51-6346, CNRS PICS  2530 and 3059,
RTN contracts MRTN-CT-2004-005104 and
MRTN-CT-2004-503369 and by a European Union Excellence Grant 
MEXT-CT-2003-509661.

\appendix

\section{Appendix: WKB approximation, useful formulae}
\label{appendix}
\renewcommand{\theequation}{A.\arabic{equation}}
\setcounter{equation}{0}

This appendix collects the results and notation of \cite{Russo:1998by}
concerning the WKB approximation. They have been used in section
\ref{sect:estim}. Consider a differential equation:
\beq
\partial_\s (f(\s) \partial_\s \phi)+ (M^2 h(\s) + p(\s))\phi=0
\eeq
where the functions $f$,$h$,$p$ behave near $\s \to 0$, $\s \to \infty$ as:
\bear
f \approx f_1 \s^{s_1}\,,\qquad
h \approx h_1 \s^{s_2}\,,\qquad
p \approx p_1 \s^{s_3}\,,\qquad
(\s\to 0)\rc
f \approx f_2 \s^{r_1}\,,\qquad
h \approx h_2 \s^{r_2}\,,\qquad
p \approx p_2 \s^{r_3}\,,\qquad
(\s\to \infty).
\eear
The WKB approximation is consistent provided $s_2-s_1+2$; 
$r_1-r_2-2$ are positive numbers and $s_3-s_1+2$; $r_1 -r_3 -2$ are positive or zero.
Define:
\beq
\a_1 = s_2 -s_1 +2 \,,\qquad \b_1  = r_1-r_2-2
\eeq
and
\bear
\a_2 = |s_1 -1| \qquad {\rm or} 	\qquad \a_2= \sqrt{(s_1-1)^2 -4 \frac{p_1}{f_1}}\qquad
({\rm if} \ \ s_3-s_1+2=0)  \rc
\b_2 = |r_1 -1| \qquad {\rm or} 	\qquad \b_2= \sqrt{(r_1-1)^2 -4 \frac{p_2}{f_2}}\qquad
({\rm if}\ \ r_1-r_3-2=0)
\eear
The masses are approximated by:
\beq
M^2 = \frac{\pi^2}{\xi^2} n (n-1+ \frac{\a_2}{\a_1} + \frac{\b_2}{\b_1}) + {\cal O}(n^0)\,,
\quad n\geq 1
\label{mwkb}
\eeq
with:
\beq
\xi= \int_0^\infty d\s \sqrt{\frac{h}{f}}
\label{xidef}
\eeq

\end{document}